# A Study on Internet of Things in Women and Children Healthcare

## Nishargo Nigar

*IEEE Member*

*nishargo@ieee.org, https://orcid.org/0000-0003-2639-7213*

## Abstract

Individual entities are being connected every day with the advancement of Internet of Things (IoT). IoT contains various application domains and healthcare is one of them indeed. It is receiving a lot of attention recently because of its seamless integration with electronic health (eHealth) and telemedicine. IoT has the capability of collecting patient data incessantly which surely helps in preventive care. Doctors can diagnose their patients early to avoid complications and they can suggest further modifications if needed. As the whole process is automated, risk of errors is reduced. Administrative paperwork and data entry tasks will be automated due to tracking and connectivity. As a result, healthcare providers can engage themselves more in patient care. In traditional healthcare services, an individual used to have access to minimal insights into his own health. Hence, they were less conscious about themselves and depended wholly on the healthcare facilities for unfortunate events. But they can track their vitals, activities and fitness with the aid of connected devices now. Furthermore, they can suggest their preferred user interfaces. This paper describes several methods, practices and prototypes regarding IoT in the field of healthcare for women and children.

## Keywords

Internet of Things

Healthcare,

eHealth,

mHealth

## 1. Introduction

The usage of IoT in healthcare services has been increased to a great extent across a range of IoT use cases. As human lives depend on healthcare, it is certainly a crucial area. Digitizing process is quite challenging for the researchers. They are constantly creating new and novel solutions on digital medical services. IoT is playing a major role in this sector. According to Cisco, about 50 billion devices will be connected to internet by the year 2020. We can trace IoT's success in the use of medical device integration, telemonitoring, smart sensors, glucose monitors, medication dispensers, activity trackers and wearable biometric sensors. Special applications like smart pills, smart beds, personal healthcare, smart home care and Real-Time Health Systems (RTHS) are latest additions to digital healthcare. Remote patient monitoring is a new way to introduce smart healthcare to mass people [1].

In underdeveloped and developing countries, women and children healthcare is an important agenda of government because of high maternal and children mortality rate. 44% decline in global maternal mortality ratio is observed between 1990 and 2015 according to Sustainable Development Goal (SDG)

Report in 2016. World Health Organization (WHO) highly puts an emphasis on the quality of antenatal care [2]. eHealth and mobile health (mHealth) system developers are designing specific healthcare applications for women and children to create awareness regarding malnutrition, obesity and mortality issues. Children healthcare faces a lot of troubles because of various factors like insurance coverage, limited resources and insufficient medicinal resources [3]. In order to polish the current healthcare facilities available to women and children, their active participation rate is mandatory. Attributes of a healthy lifestyle consists of several activities, proper knowledge of nutrition, appropriate diet plan and awareness of risky diseases. Implementation of such healthcare applications [4] [5] assists in active involvement of users. In addition, the mHealth and eHealth solution developers come to learn about user interactions, faults present in traditional systems and useful feedback.

User experience must be taken into consideration while making an innovative approach. The dynamic functionalities of IoT is making it possible to ensure user experience and it has been possible after a lot of research on security, protocols, embedded devices, applications, interoperability and network architectures. To deploy the technology, a lot of countries are applying relevant policies and guidelines [6]. Women and children healthcare is considered as a sensitive case because of its vulnerable and ill-fated statistics. It is necessary to implement the most appropriate and efficient systems to uphold their health status. But the pressure of rising costs in healthcare systems can decelerate the whole process. It is a matter of elation that various software and connected devices are available to mitigate some financial problems that the healthcare providers are facing at present. This is being possible by the recent advancements of IoT.

In this paper, we made an effort to highlight the digital healthcare solutions based on IoT available for women and children and how they are reshaping the current scenario of conventional healthcare management systems. For the most part, an extensive set of prototypes available for women and children are demonstrated throughout our paper. We also described the challenges and ongoing trends of present digital healthcare systems for the target users.

## 2. Applications & Services

Applications of IoT are rising rapidly due to our needs and demands. Radio frequency Identification (RFID), machine-to-machine (M2M) and Near-field communication (NFC) are being applied and creating a revolution in the field of IoT. Although the concept of M2M might sound complex, the idea behind the execution is pretty simple. M2M networks function as Wide Area Network (WAN) or Local Area Network (LAN) but primarily used for permitting controls, sensors and machines to communicate among each other. The devices feed information acquired to other devices present in the same network. Such system allows a person to assess what is happening in the entire network environment and issue necessary instructions to member devices.

Various researches have been conducted until the present relating to IoT targeting women and children. We have created an exclusive category of works and prototypes related to IoT in women and children healthcare in figure 1. From time to time, researchers have built smart devices, proposed new methods or suggested enhanced systems for the welfare of the target group. We have also discussed about their drawbacks so that in future, researches can be conducted solving the issues.

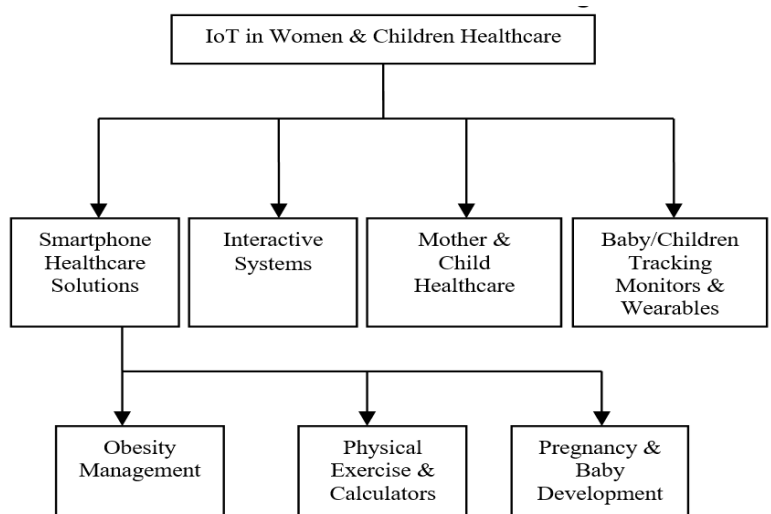

Figure 1. IoT applications and services targeting women and children

## 2.1. Smartphone Healthcare Solutions

Nowadays, smartphones are capable of doing more than making phone calls or sending texts. They are equipped with Global Positioning System (GPS), Wireless Fidelity (Wi-Fi), NFC, voice recognition and sensors which make them so productive [8]. Today's smartphones consist of almost 14 sensors that generate raw data of location, motion and surrounding environment. With the aid of micro-electromechanical systems (MEMS), this revolution has been possible. MEMS structure is composed of a number of four elements: microsensors, microactuators, microstructures and microelectronics (shown in fig. 2). Microsensors and microactuators are classified as transducers that exchange energy from one type to another. The device turns a mechanical signal into an electrical one when it comes to microsensors. But the main purpose of MEMS is to combining such tiny structures into a regular silicon substrate with integrated circuits. Micromechanical elements are fabricated with micromachining process where sections of silicon wafer are incised or new layers are added to obtain electromechanical devices.

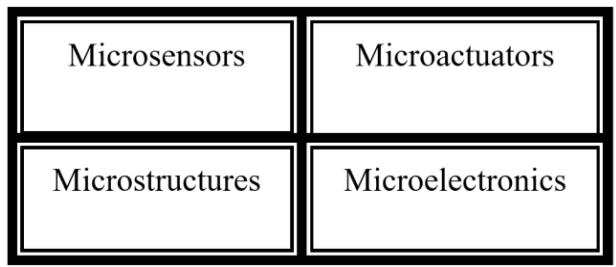

Figure 2. Components of micro-electromechanical systems (MEMS)

The major sensors in a smartphone include: magnetometer, GPS, gyroscope, accelerometer, barometer, proximity sensors and so on. Magnetometer is actually known as a compass and with GPS, it can detect one's current location. In cell phones, GPS receivers communicate with available units from an array of 30 global positioning satellites within the system. Trilateration (process of detecting absolute location points) is done with built-in receiver from minimum three satellites and the receiver. GPS can measure one's location by an equation based on the GPS receiver of cell phone and intersection point of overlying spheres gained from satellites. On the other hand, accelerometer in a smartphone is comprised of a circuit with seismic mass which means it is made from Silicon. In fact, it is a circuit with the basis of MEMS. The circuit senses a force of acceleration due to tilting or movement of the phone.

Several smartphone based healthcare solutions were created previously targeting women and children. They are of different subjects and offer distinct services towards healthcare management of an individual.

They are mostly based on nutrition facts, obesity management, exercise, diet plan, calorie counter, step counter, period tracker and pregnancy management.

### 2.1.1. Obesity Management

Concept of IoT is utilized to spread knowledge of various food items and obesity management. In the past, Automatic Identification and Data Capture Techniques (AIDC) has been used in mHealth platform to create health awareness among children. In AIDC, three major elements are necessary (fig. 3); the first one is a data encoder. A code represents symbols that are generally alphanumeric characters. While encoding data, characters are turned into machine readable format. A tag including encoded is attached to an item that needs to be detected. A scanner assists in reading encoded data and converts them to an electrical signal. Finally, a data decoder converts the signal to digital data and returns to the initial alphanumerical characters. RFID uses radio waves to implement AIDC technique. An RFID tag holds an antenna and integrated circuit to send data to RFID reader. Reader transforms the radio waves to a usable form and the data gained from tags are transmitted with the aid of a communication interface to a particular host system so that data can be recorded in the database and examined if necessary.

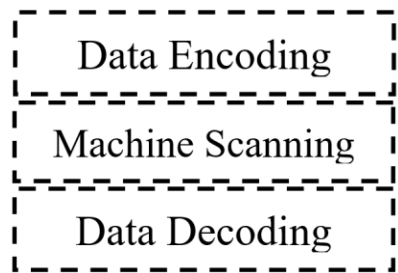

Figure 3. Three Steps of AIDC Technique

In the system, notifications are delivered to the tutors after scanning the Quick Response (QR) code or RFID tag stored in the food item. This tag includes essential information about the food like its nutritive elements: sugar, saturated fat, calories and sodium. The paper followed two implementation approaches; one with using RFID and another one is QR code. RFID tag is readable with a mobile device that is NFC-enabled. But RFID has a disadvantage; while utilizing RFID tags, maximum users need to alternate their handheld devices to an NFC-enabled version. That is why Bluetooth can be a great alternative. On the other hand, QR codes are easily decoded by a scanner and a camera of a mobile phone [9]. The downside of the research is- it is based on a really old technology. For example, it works with HTC Magic (Operating System: Android 1.6 Donut) and Nokia N-95 (Operating System: Symbian OS v9.2, S60 $3^{rd}$ edition). In addition, iOS platform can also be tested for such mHealth technology.

At this time, plenty of eHealth services are accessible to general people in order to maintain proper health condition. Mobile applications can be very constructive when it comes to managing healthy lifestyles. The research works reported in [10] [11] seem to have positive outcomes concerning the subject. The best thing about mobile applications is they are developed in a personalized manner for an individual. For example, data are being added by a user everyday and the user can monitor the activities anytime in future.

It has been proved that games are useful in case of a child's memory and critical thinking skills. They can also have positive impacts such as enhancing analytical thinking skills, conceptual learning capabilities, problem solving skills, creativity, linguistic improvement and so on. Designing a game or mobile application for a child consists of several important components like prototyping, design, production, testing, fixing bugs, maintenance etc. shown in figure 4.

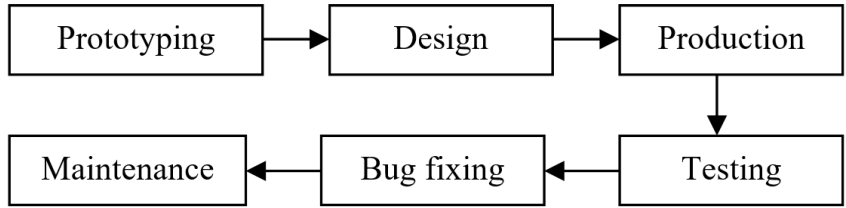

Figure 4.Development process of mobile applications and games

In [12], the authors proposed a mobile application as an informative tool to prevent increased obesity rates. The application is based on IoT and it includes remote capturing, tracking food intake and continuous monitoring system for children. The application creates awareness among children regarding healthcare and also explores nutrition benefits. The mobile application is certainly great for preventing obesity rates, but it could suggest a methodology for children suffering from malnutrition or undernutrition as well. A mHealth platform [13] using transfer learning and IoT has been proposed to improve children's health consciousness and behavior. In the paper, a balanced and categorized food chart according to children's meal time and nutrition has been prepared. The task is done using transfer learning and You only look once (YOLO) algorithm. Transfer learning has two approaches; a. Pre-trained model approach and b. Develop model approach where the former one is chosen for the platform. In pre-trained approach, a source model is chosen at first and then the model is reused as the initial point for other model on another task. Tuning might be necessary to modify data of input-output pair. In YOLO algorithm, a neural network is applied to a full image. After that, the image is partitioned into few regions. Thus, bounding boxes for each region are found. Height, weight, (x, y) coordinates and confidence scores are received from the bounding box. To find out confidence scores for the boxes, the following equation is used;

$$\Pr(Class_i | Object) * \Pr(Object) * IOU = \Pr(Class_i) * IOU \quad (1)$$

where, $IOU$ = Intersection over union.

In the module, two subsystems are present for child user and health instructor. They are both connected for communication purpose. The platform holds a practical value for children as whenever a child user is capturing a photo of the meal he is going to take through the camera, he is receiving the nutrition value through the module. Although the platform is beneficial for children, it lacks suggestion on how to avoid junk foods. *Health Attack* [14] is a mobile application based on iPhone platform targeting afro-american children who are aged between 7 and 11 years old focused on informing them about nutritional value of diverse food items. There is a total of 7 different nutrients; water, vitamins, fats, proteins, minerals, carbohydrates and fiber. Children should gain the idea of them when they are developing their mental and physical abilities. The downside of the app is it is static in nature and more functionalities regarding practical approaches should be added.

### 2.1.2. Physical Exercise & Calculators

The works presented in [15] [16] inspired children and teenagers to perform physical activities more often. Mobile games have been created with the purpose of fulfilling the requirements. Women can utilize the menstrual cycle calculator apps like Maya Apa [17], which is a popular app in Bangladesh. The app allows users to ask questions anonymously so that they never feel uncomfortable to know about any sensitive health information. But a big disadvantage is, as the user can enter the system without providing any personal information, many unpleasant posts appear in the app anonymously written by ill-natured users. There are many apps online which present integrated Body Mass Index (BMI) and Total Daily Energy Expenditure (TDEE) calculators. To determine one's BMI and TDEE, following equations are followed:

$$BMI = \frac{Weight\ (lbs)}{Height(in)^2} * 703 \quad (2)$$

$$\text{Or, } BMI = \frac{Weight(kg)}{Height(m) * Height(m)}$$

TDEE formula applied for women:

1. Measure Basal Metabolic Rate (BMR) using;

$$BMR = (height\ in\ cm * 6.25) + (weight\ in\ kg * 9.99) - (age * 4.92) - 161 \quad (3)$$

2. To calculate TDEE;

$$TDEE = BMR * activity\ level \quad (4)$$

For sedentary lifestyle; $TDEE = BMR * 1.1 \quad (5)$

Lightly active lifestyle; $TDEE = BMR * 1.275 \quad (6)$

Moderately active lifestyle; $TDEE = BMR * 1.35 \quad (7)$

Very active lifestyle; $TDEE = BMR * 1.525 \quad (8)$

After taking inputs from user, the calculators show the results and recommend a standard score and suggest how to achieve the goal. The shortcoming of these calculators is that a new kind of calculator is available now which is more accurate and simpler than the previous ones. This calculator is based on relative fat mass index (RFM). In this case, one only requires a tape measure rather than scales. To determine RFM of a woman;

$$76 - \left(20 * \frac{height}{waist\ circumference}\right) = RFM(women) \quad (9)$$

On the other hand, trending smartphone applications consist of nutrition facts for children that are available on Apple App Store and Google Play Store. These facts are taught in an entertaining manner. Recently, edutainment has taken the concept of education to next level. These applications strongly motivate a child's access to the understanding of learning, health and nutrition through games.

### 2.1.3. Pregnancy & Baby Development

For mothers, apps pertaining to pregnancy are available containing tips of morning sickness, prenatal vitamins and workout tips to avoid pregnancy weight gain and so on. They also proffer features like pregnancy calendar, baby growth detector, baby due date calculator and ovulation calculator. An example is *Pregnancy Tracker & Baby App* [18]. Pregnancy is a long process where a mother must be careful about her health and should take nutrient-rich foods. Such apps are becoming very popular among pregnant mothers because of their wide-ranging set of features. The biggest drawback of these apps is it does not include any real time services or features. A pregnant woman might need an urgent support during the months of pregnancy. But such features are still not available yet.

We have conducted an exclusive survey with 23 women who uses mobile applications for various healthcare services. Among these 23 users, 16 women use Android platform, 5 women use iOS platform and 2 of them uses Windows platform. In addition, the types of health apps and the problems they face using such apps are demonstrated in figure 5 and 6 respectively. Most of the women uses multiple apps to maintain their health. The survey illustrates 14 users use health apps for exercise and fitness, 9 of them use health apps for counting calories, 12 of them use such apps for maintaining diet and weight, 10 of the users keeps track of their period or menstrual cycle, 3 of them use them for pregnancy purpose, 8 of them check blood pressure with the aid of such apps, 4 of them use apps for sleep and meditation purpose and 5 of them also use the health apps for other tasks.

In response to the difficulties, 9 of the women users stated that such health apps often contain bad user interface, 3 of them mentioned minor bug issues, 8 of them said they hold no practical values, 12 of them pointed out that they only have static features and 4 of them stated they need to be updated more often.

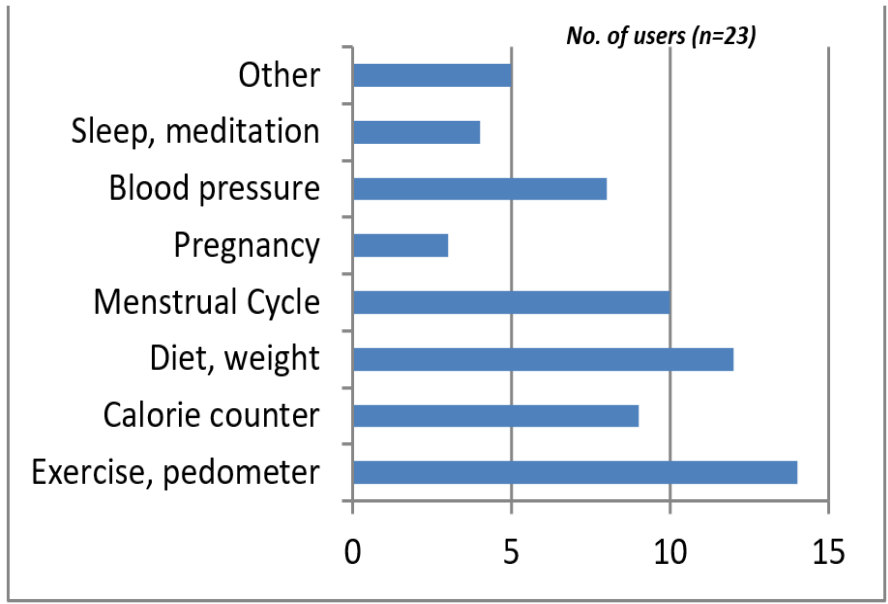

Figure 5.Health apps and their users (women)

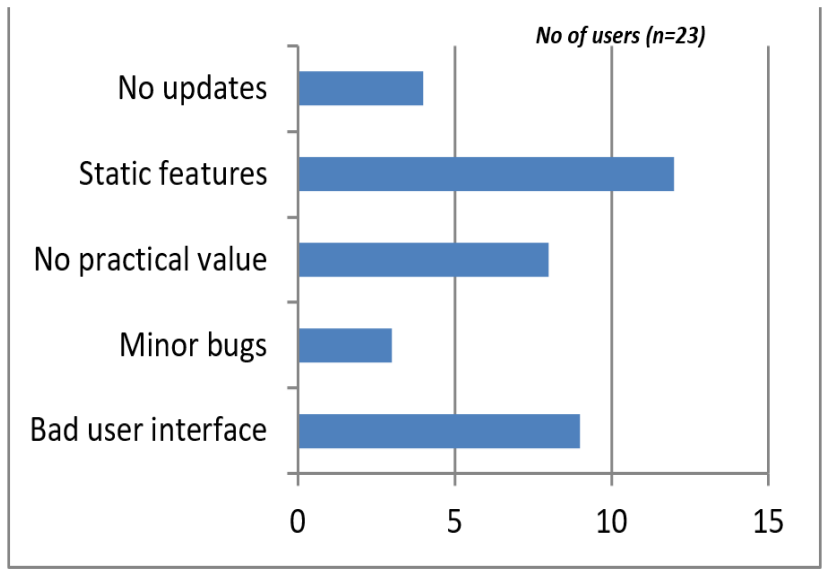

Figure 6.Health apps and their problems (women)

We believe major changes can be applied for the welfare of these health apps and their users. According to our survey, there are lots of opportunities in health apps, as well as challenges to overcome.

## 2.2. Interactive Systems

An IoT enabled interactive totem [19] was developed in a paediatric ward of a hospital to improvise hospitalization experience for children. With the help of the totem, children can get friendly with the unknown environment surrounded by doctors, nurses and tons of medical equipment. The totem involves the feature of self-ordering meals where a child is recognized by the system and his favorite meal is suggested. It contains webcam, stereo speakers, 26'' LCD touch screen and internet. The system mentioned in the paper lacks important features like emergency support, food detection and nutrition facts. In addition, what happens after a child patient is discharged is not discussed.

Usually when applying recognition feature in an IoT system, it is also really important to activate triggers for each actions. For such purposes, various tools are available now. Combining ThingSpeak and If this, then that (IFTTT) platforms can result in a number of useful services for children and women. There are also other software tools that have potential to generate improved solutions for this target group. Table 1 discusses the collection of software tools that can be used for this purpose.

**Table 1.** Software tools for IoT environment

| Software tool | Function/Purpose |
|---|---|
| Windows IoT | Operating system |
| AllJoyn | Open source software framework |
| Node-Red | Flow-Chart based development tool |
| openHab | Open source automation software for home |
| MindSphere | "IoT operating system"/open cloud platform |
| IOTivity | Open source software framework |
| Android Things | Android-based embedded operating system |

An intelligent children healthcare system is developed in [20] where the system tracks records, monitors progress and promotes nutrition through mHealth and edutainment features. After analyzing the data, the results were recorded with big data analytics tool called Hadoop. Finally, they were compared with Logistic Regression and Naïve Bayes algorithms where Logistic Regression performed better than the other. The system only detects if a child is in good health condition or not. It lacks recommendation or suggestion feature based on a child's current health state. In [21], a smart system was designed by the researchers with the help of IoT and MapReduce framework of Hadoop to detect anxiety disorders, disruptive behavior disorders and Attention-deficit/hyperactivity disorder (ADHD). MapReduce framework has two classes; a. Mapper and b. Reducer. Mapper class reads data blocks and create key-value pairs as outputs. The output is taken as an input to reducer class and it combines key-value pairs to smaller group of key-value pairs. To classify the child behavior, C4.5 decision tree algorithm has been used. In [22], the authors proposed an intelligent system with a smart nutrition card that holds a child's daily activity, calorie intake, emotional state and performance after collecting the data through smartphone sensors, recommendation methods and artificial intelligence. The interactive system collects the child's attendance rate, emotional and physical state, BMI & TDEE updates, activities, nutrition info and real time health issues. The negative aspect of the research is that the smart nutrition card needs to be practically implemented with actual users in a hospital environment.

## 2.3. Mother & Child Healthcare

To reduce infant and maternal mortality rate, Mother and Child Tracking System (MCTS) was developed in India [23]. The system engages caregivers with patients through web-based application and SMS. A kind of similar prototype was also done in Nigeria [24]. In Indonesia, a multiplatform system [25] was designed for mother and children care to detect diseases early through remote monitoring service. The system includes multiple sensors, mobile application and a portal. All of it is maintained by a cloud so that caregivers and patients can obtain easy access. The system offers positive healthcare experience in terms of mother and children. The biggest drawback of such multiplatform systems is security vulnerability due to cloud management system depicted in figure 7. A lot of devices are using cloud services for various purposes and it is indisputably unsafe because of hackers. These smart devices and applications need strong cryptographic support. Otherwise, the private information can be breached unintentionally.

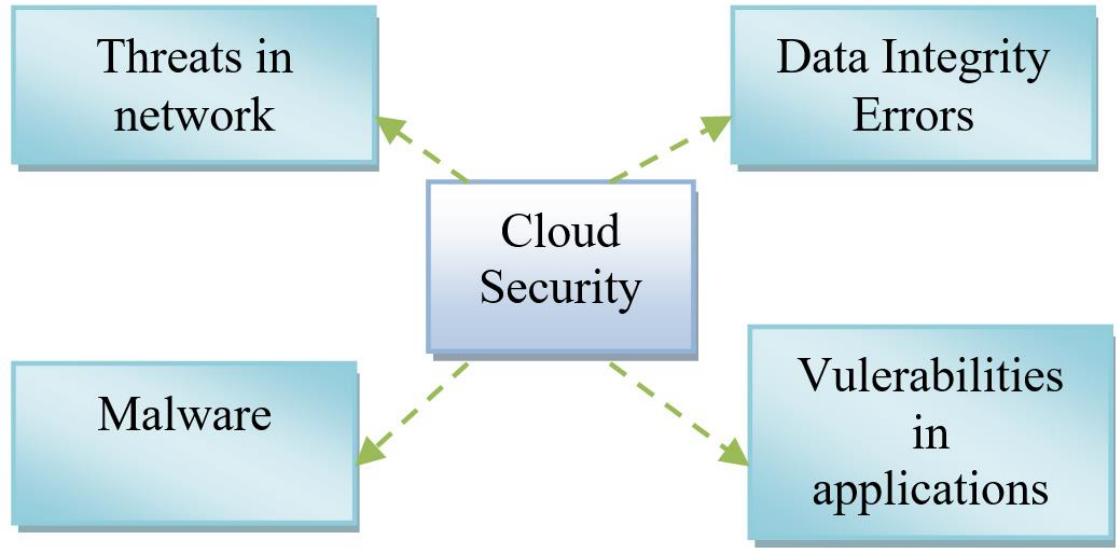

Figure 7. Cloud Security Issues

## 2.4. Baby/Children Tracking Monitors & Wearables

A real-time location infant monitoring system has been developed in [26] with wristband active RFID tags. Using this, the newborn's important data like current location and body temperature can be tracked without any difficulty. The baby's location is recognized by decision tree classifiers. Although the infant monitoring system is quite constructive, it may contain potential health hazards especially for children. A safety system for school going children [27] on a bus has been designed so that the parents can make sure that their children are on the right track. This research could be quite simpler if they could apply GPS for the bus riders into their system. A child safety wearable device [28] using Short Message Service (SMS) and Global System for Mobile (GSM) communication is created for parents to know about their child's current location and temperature. The system utilizes Arduino Uno, a visual distress signal known as SOS light, temperature sensor, Ultraviolet (UV) sensor, GPS sensor, alarm buzzer, GSM shield and base shield. In [29], a wearable device called *Raksh* was built. The aim of the paper was to fight against pneumonia from which children might suffer a lot. Special IoT based wearable devices are presented in [30] [31] [32] [33] targeting women and girl children for ensuring their safety in adverse and dangerous situations. These papers reflect how girls can be protected in unsafe situations with the help of smart devices. But unfortunately, smart gadgets like wearables hold major privacy concerns and unwanted health issues due to high radiation levels.

Table 2 discusses the collection of software tools that can be used for this purpose.

**Table 2.** Comparative analysis among used methods or platforms for women & children

| Used methods/platforms | Merits | Demerits |
|---|---|---|
| RFID | Real-time tracking, efficient, reduces errors | Sometimes alternating to NFC enabled versions is mandatory |
| Games (children) | Holds constructive and practical values, a great source of edutainment | Traditional approach, common ideas |
| Mobile Applications (children) | Simple and easy | Static features, raw data only |
| Calculators (women) | Helps to keep track | Accuracy is questionable |
| Interactive Systems | Face recognition, personalized choices | Lacks emergency support |
| Mobile Applications related to pregnancy and baby development | Important tips, calendars | No real-time services available, no emergency support |
| Multiplatform Systems | Better healthcare experience, getting connected to caregivers | Vulnerable cloud security issues |

In addition, we have applied Analytic Hierarchy Process (AHP) in order to find out the desired user interface for such systems. Common user interface types include form, menu, graphical user interface (GUI), command line and natural language. We have considered three major options: form based, menu driven and GUI in case of comfort in usability.

A pair-wise comparison is a must to select preferences for criteria and choices. For each matrix that includes pair-wise elements,

$$\begin{bmatrix} C_{11} & C_{12} & C_{13} \\ C_{21} & C_{22} & C_{23} \\ C_{31} & C_{32} & C_{33} \end{bmatrix}$$

In the table shown below (Table 3), intensity judgment is done with pairwise comparison matrix. A = Form based, B = GUI and C = menu driven where GUI is more preferable than form based and also more preferable than menu driven approach. So,

**Table 3.** Comparison of matrix of Form based, GUI and menu driven approach

| Choosing UI | A | B | C |
|---|---|---|---|
| A | 1 | 1/3 | 2 |
| B | 3 | 1 | 5 |
| C | ½ | 1/5 | 1 |

Now, the values are summed up in each of the column of pair-wise matrix in (10),

$$C_{ij} = \sum_{i=1}^{n} C_{ij} \quad (10)$$

Thus, to find out the normalized matrix, addition of columns must be done. The process is shown in table 4 and 5.

**Table 4.** Addition of columns

| Choosing UI | A | B | C |
|---|---|---|---|
| A | 1.000 | 0.333 | 2.000 |
| B | 3.000 | 1.000 | 5.000 |
| C | 0.500 | 0.200 | 1.000 |
| Total | 4.500 | 1.533 | 8.000 |

**Table 5.** Normalized matrix & sum of rows

| Choosing UI | A | B | C | Total |
|---|---|---|---|---|
| A | 0.22 | 0.217 | 0.25 | 0.687 |
| B | 0.67 | 0.652 | 0.625 | 1.947 |
| C | 0.111 | 0.130 | 0.125 | 0.366 |

To generate a normalized pair-wise matrix, each element should be divided by its column in the matrix like shown in equation (11).

$$X_{ij} = \frac{C_{ij}}{\sum_{i=1}^{n} C_{ij}} \begin{bmatrix} X_{11} & X_{12} & X_{13} \\ X_{21} & X_{22} & X_{23} \\ X_{31} & X_{32} & X_{33} \end{bmatrix} \quad (11)$$

To create the weighted matrix, equation (12) is used where n = no. of criteria,

$$W_{ij} = \frac{\sum_{j=1}^{n} X_{ij}}{n} \begin{bmatrix} W_{11} \\ W_{12} \\ W_{13} \end{bmatrix} \quad (12)$$

Normalizing sum of rows produce,

$$\begin{bmatrix} 0.687/3 \\ 1.947/3 \\ 0.366/3 \end{bmatrix} = \begin{bmatrix} 0.229 \\ 0.649 \\ 0.122 \end{bmatrix} \quad (13)$$

According to the priority matrix, GUI has the maximum priority. Finally we need to measure the consistency ratio (CR) which should be <=10% in accordance with Saaty, who developed the original method. To get consistency ratio, we have to calculate consistency index (CI) at first which is based on equation (14),

$$\begin{bmatrix} C_{11} & C_{12} & C_{13} \\ C_{21} & C_{22} & C_{23} \\ C_{31} & C_{32} & C_{33} \end{bmatrix} * \begin{bmatrix} W_{11} \\ W_{21} \\ W_{31} \end{bmatrix} = \begin{bmatrix} Cv_{11} \\ Cv_{21} \\ Cv_{31} \end{bmatrix} \quad (14)$$

After that, division of weighted sum vector is executed with criterion weight.

$$Cv_{11} = \frac{1}{W_{11}}[C_{11}W_{11} + C_{12}W_{21} \; C_{13}W_{21}] \quad (15)$$

$$Cv_{21} = \frac{1}{W_{21}}[C_{21}W_{11} + C_{22}W_{21} \; C_{23}W_{31}] \quad (16)$$

$$Cv_{31} = \frac{1}{W_{31}}[C_{31}W_{11} + C_{32}W_{21} \; C_{33}W_{31}] \quad (17)$$

To determine $\lambda$, average value of consistency vector is measured. CI is used to perceive deviation.

$$\lambda = \sum_{i=1}^{n} Cv_{ij} \quad (18)$$

$$CI = \frac{\lambda - n}{n - 1} \quad (19)$$

Therefore, we follow the aforementioned steps to get our desired values,

$$\begin{bmatrix} 1 & 1/3 & 2 \\ 3 & 1 & 5 \\ 1/2 & 1/5 & 1 \end{bmatrix} * \begin{bmatrix} 0.229 \\ 0.649 \\ 0.122 \end{bmatrix} = \begin{bmatrix} 0.69 \\ 1.95 \\ 0.37 \end{bmatrix}$$

$$\begin{bmatrix} 0.69/0.229 \\ 1.95/0.649 \\ 0.37/0.122 \end{bmatrix} = \begin{bmatrix} 3.01 \\ 3.00 \\ 3.03 \end{bmatrix}$$

$$\lambda = \frac{3.01 + 3.00 + 3.03}{3} = 3.01$$

$$CI = \frac{3.01 - 3}{3 - 1} = 0.005$$

Finally, to obtain CR, equation (20) is used;

$$CR = \frac{Consistency\ Index(CI)}{Random\ Index\ (RI)} \quad (20)$$

$$CR = \frac{0.005}{0.90} = 0.0055\ (acceptable)$$

Therefore, our calculation proves that GUI is the most preferred option as the user interface than form based or menu driven interfaces. To make solutions better, one should put an emphasis on generating creative, simple and comfortable GUI for children and women users.

## 3. Current Trends & Challenges

The annual growth rate of internet of medical things is estimated to reach around $72 billion by the year 2021. IoT is being adopted by the healthcare community for its extensive features and potentials. Implantable devices, wearables and monitors will carry on sending real time data to hospitals. For mothers, a smart toy called 'Teddy the Guardian' can be very helpful as it tests baby's temperature and heart rate. Not only that, the smart stuffed toy checks the oxygen saturation level when it receives a hug from the baby. Care-plan-specific mobile applications are brought into play to decrease readmission rates. The concept of smart bag (with microcontroller ATmega16) and smart shirt (enabled with NFC chip) for children are on the rise. Virtual assistants and mobile health applications have become popular already. Healthcare in IoT will sooner or later result in better patient experience, improved outcomes, superior disease management and lower treatment costs.

Devices that are connected to internet can come in a variety of appearances. Data can be collected from temperature monitors, wearables, healthcare apps or interactive systems. Most of these calculations need a follow-up communication with a doctor or a professional in the area. It generates an opportunity for smart devices to distribute precious data. IoT is surely making a difference in the industry of healthcare. Data management in hospitals has been improved in an overwhelming manner caused by IoT devices like monitors and scanners. Diseases like diabetes are being detected by such digital healthcare services. A transformation is taking place in healthcare industry and business organizations should come forward to invest in latest technologies with the aim to enhance medical care. They are already grabbing the opportunity and creating personalized solutions for individuals and particular entities. Smart hospitals are allowing RFID enabled pharmacy inventory system, automated drug delivery, smart patient flow, real-time location data sharing and smart consoles with the intention of treating patients more precisely. Researchers and developers across the earth are exploring vivid solutions that complement previous services by activating the possibilities of IoT.

IoT devices are not restricted to a small network. The devices receive and send personal health information (PHI). Such information must be kept protected. IoT health monitoring devices are directly connected to the internet. As a result, privacy is the most vulnerable security issue of smart healthcare services. Once upon a time, we only considered the privacy issues of our personal computers. Then the smartphones joined the market. And finally we have the next big thing here, which happens to connect all the gadgets in a single platform – wearables, home appliances, car, phone, remote control and so on.

## 4. Conclusion

Numerous applications and services have been developed yet for the welfare of women and children. Active participation of this target group is mandatory to ensure the success of the eHealth, mHealth and IoT services built for them. In the paper, we discussed about their possibilities and benefits, as well as the challenges to overcome. The paper demonstrates a new perspective to make use of technology for women and children healthcare. IoT has promising economic and technological prospect in spite of its security issues. We believe that IoT will change the traditional healthcare system eventually for the greater good, especially for women and children.

## Author's Biography

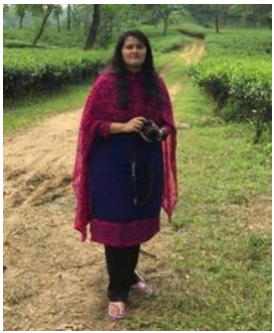

**Nishargo Nigar** holds a Bachelor of Science in Computer Science & Engineering from East Delta University, Chittagong, Bangladesh. She is an entrepreneur and she is running two IT startups currently. She is a student member of IEEE. Her research interests are Internet of Things, mHealth and machine learning.

## How to Cite